\documentclass{aastex}
\usepackage{emulateapj5}
\usepackage{apjfonts}

\shorttitle{Observations of GRB~010921}
\shortauthors{Park et al.}

\begin{document}


\title{LOTIS, Super-LOTIS, SDSS and Tautenburg Observations of GRB 010921}


\author{
H. S. Park\altaffilmark{1},
G. G. Williams\altaffilmark{2},
D. H. Hartmann\altaffilmark{3},
D. Q. Lamb\altaffilmark{4,16}, 
B. C. Lee\altaffilmark{5}, 
D. L. Tucker\altaffilmark{5}, 
S. Klose\altaffilmark{6},
B. Stecklum\altaffilmark{6},
A. Henden\altaffilmark{13},
J. Adelman\altaffilmark{5},
S. D. Barthelmy\altaffilmark{7},
J. W. Briggs\altaffilmark{17},
J. Brinkmann\altaffilmark{12},
B. Chen\altaffilmark{10,11},
T. Cline\altaffilmark{7},
I. Csabai\altaffilmark{10,18},
N. Gehrels\altaffilmark{7},
M. Harvanek\altaffilmark{12},
G. S. Hennessy\altaffilmark{19},
K. Hurley\altaffilmark{8},
\v{Z}eljko Ivezi\'{c}\altaffilmark{14}, 
S. Kent\altaffilmark{5},
S. J. Kleinman\altaffilmark{12}, 
J. Krzesinski\altaffilmark{12,21},
K. Lindsay\altaffilmark{3}, 
D. Long\altaffilmark{12}, 
R. Nemiroff\altaffilmark{9},
E. H. Neilsen\altaffilmark{5},
A. Nitta\altaffilmark{12},
H. J. Newberg\altaffilmark{20},
P. R. Newman\altaffilmark{12}, 
D. Perez\altaffilmark{9},
W. Periera\altaffilmark{9},
D. P. Schneider\altaffilmark{15},
S. A. Snedden\altaffilmark{12},
C. Stoughton\altaffilmark{5},
D. E. Vanden Berk\altaffilmark{5}, 
D. York\altaffilmark{4,16},
K. Ziock\altaffilmark{1}
}

\altaffiltext{1}{Lawrence Livermore National Laboratory, 7000 East
Ave., Livermore, CA 94550.}

\altaffiltext{2}{Steward Observatory, University of Arizona, Tucson, AZ 
85721.}

\altaffiltext{3}{Department of Physics \& Astronomy, Clemson University,
Clemson, SC 29634-0978.}

\altaffiltext{4}{Department of Astronomy and Astrophysics, University
of Chicago, 5640 South Ellis Avenue, Chicago, IL 60637.}
 
\altaffiltext{5}{Experimental Astrophysics Group, Fermi National
 Accelerator Laboratory, P.O. Box 500, Batavia, IL 60510.}

\altaffiltext{6}{Th\"uringer Landessternwarte Tautenburg, 
07778 Tautenburg, Germany}
  
\altaffiltext{7}{NASA/Goddard Space Flight Center, Greenbelt, MD 20771}

\altaffiltext{8}{Space Sciences Laboratory, University of California, 
Berkeley, CA 94720}

\altaffiltext{9}{Department of Physics, Michigan Technological University, 
Houghton, MI 49931}

\altaffiltext{10}{Department of Physics and Astronomy, Johns Hopkins
University, 3701 San Martin Drive, Baltimore, MD 21218.}

\altaffiltext{11}{XMM Science Operation Center, European Space 
Agency - Vilspa, Villafranca del Castillo, Apartado 50727 - 28080 
Madrid, Spain.}

\altaffiltext{12}{Apache Point Observatory, P.O. Box 59, Sunspot, NM
88349-0059.}
 
\altaffiltext{13}{Universities Space Research Association / U. S. Naval
Observatory, Flagstaff Station, P. O. Box 1149, Flagstaff, AZ 86002-1149.}

\altaffiltext{14}{Princeton University Observatory, Peyton Hall, 
Princeton, NJ 08544-1001.}

\altaffiltext{15}{Astronomy and Astrophysics Department, Pennsylvania 
State University, 525 Davey Laboratory, University Park, PA 16802.}

\altaffiltext{16}{Enrico Fermi Institute, University of Chicago, 
5640 South Ellis Avenue, Chicago, IL 60637.}

\altaffiltext{17}{Yerkes Observatory, University of Chicago, 
373 West Geneva Street, Williams Bay, WI 53191.}

\altaffiltext{18}{Department of Physics of Complex Systems, 
E\"otv\"os University, P\'azm\'any P\'eter s\'et\'any 1, 
H-1518, Budapest, Hungary.}

\altaffiltext{19}{U.S. Naval Observatory, 3450 Massachusetts Ave., 
NW, Washington, DC  20392-5420.}

\altaffiltext{20}{Physics Department, Rensselaer Polytechnic 
Institute, SC1C25, Troy, NY 12180.}

\altaffiltext{21}{Mt. Suhora Observatory, Cracow Pedagogical University,
ul. Podchorazych 2, 30-084 Cracow, Poland.}



\begin{abstract}
We present multi-instrument optical observations of the High Energy 
Transient Explorer (HETE-2)/Interplanetary Network (IPN) error box 
of GRB~010921. This event
was the first gamma ray burst (GRB) localized by
HETE-2 which has resulted in the detection of an optical afterglow.
In this paper we report the earliest known observations of the
GRB010921 field, taken with the 0.11-m Livermore Optical Transient
Imaging System (LOTIS) telescope, and the earliest known detection
of the GRB010921 optical afterglow, using the 0.5-m Sloan Digital
Sky Survey Photometric Telescope (SDSS PT). Observations with the
LOTIS telescope began during a routine sky patrol 52 minutes after
the burst. Observations were made with the SDSS PT, the 0.6-m
Super-LOTIS telescope, and the 1.34-m Tautenburg Schmidt telescope
at 21.3, 21.8, and 37.5 hours after the GRB, respectively. In addition, 
the host galaxy was observed with the USNOFS 1.0-m telescope
56 days after the burst.
We find that at later times (t > 1 day after the burst), the
optical afterglow exhibited a power-law decline with a slope of
$\alpha = 1.75 \pm 0.28$.  However, our earliest observations show
that this power-law decline can not have extended to early times
(t < 0.035 day).
\end{abstract}


\keywords{gamma rays: bursts---gamma rays: observations}

\section{Introduction}

The High Energy Transient Explorer (space.mit.edu/HETE/) 
HETE-2 is dedicated to the study of gamma-ray bursts (GRBs).
HETE-2 is currently the only GRB detector capable
of localizing and disseminating GRB coordinates in near real-time. 
Low-energy emission during and shortly after a GRB ($t \lesssim 1$~hr) 
potentially holds the key to significant progress in understanding the 
central engine of GRBs and could provide valuable clues to their progenitors
\citep{meszaros01}. 

The HETE-2 detection of GRB~010921 together with data from
the Interplanetary Network (IPN) provided the first HETE-2
localization which has resulted in the detection of an
optical afterglow. Although the afterglow was relatively 
bright, early observations of the error box failed to reveal 
any candidate afterglows because of source confusion with its
bright host galaxy ($R \sim 21.7$) \citep{price01a}. Spectroscopy
of the host galaxy performed with the Palomar 200-in
telescope four weeks after the burst indicates a redshift
of $z$~$=~0.450~\pm~0.005$ \citep{djorg01a}.

\section{Observations}

\subsection{GRB Observations by HETE-2 and IPN}

On 21 September 2001 at 05:15:50.56~UT (September 21.21934~UT)
the Fregate instrument detected
a bright GRB (Trigger\ 1761). GRB~010921 had a duration of
$\sim 12$~s in the 8-85~keV band, a peak
flux of ${\rm F_{p}} > 3\times10^{-7} {\rm erg~cm}^{-2}~{\rm s}^{-1}$
and a fluence of
$S \sim 1\times10^{-6} {\rm erg~cm}^{-2}$ \citep{ricker01a}. For a 
spatially flat FRW cosmology ($\Omega_m$ = 0.3 , $\Omega_\Lambda$ = 0.7,
and $H_0$ = 65 km s$^{-1}$ Mpc$^{-1}$) the measured redshift implies 
an equivalent isotropic energy release of $6 \times 10^{50}$~erg. 
This inferred energy would be within the range suggested by \citet{frail01} 
for the standard energy reservoir, $E_0 \sim 5 \times 10^{50}$~erg, 
suggesting the opening angle of the gamma-ray jet in GRB~010921 is modest. 

The burst was also detected in the Wide Field X-ray Monitor 
WXM~X detector but was outside
the field-of-view of the WXM~Y detector \citep{ricker01b}. 
Because the GRB was not well localized in the Y direction, a 
location was not
distributed with the real-time trigger. Approximately five
hours after the GRB, ground analysis gave 
a long narrow error box ($\sim 10\arcdeg \times \sim 20\arcmin$)
centered at 
$\alpha, \delta = 23^{\rm h}2^{\rm m}14\fs6, 44\arcdeg16\arcmin4.8\arcsec$
(J2000.0) \citep{ricker01a}. 
{\it Ulysses} and {\it BeppoSAX} also detected GRB010921. 
These detections resulted
in an IPN triangulation approximately 15~hours after the burst.
The IPN location for the burst is an annulus centered at 
$\alpha, \delta = 15^{\rm h}29^{\rm m}11\fs8, 67\arcdeg36\arcmin18.0\arcsec$ 
(J2000.0), with radius of $60\fdg003 \pm 0\fdg156 (3\sigma)$.
The combined HETE-2/IPN data resulted in an error
box with an area 
of a $310 \sq \arcmin$ box with corners at 
$\alpha_1, \delta_1 = 22^{\rm h}54^{\rm m}21\fs87, 
40\arcdeg36\arcmin25.84\arcsec$,
$\alpha_2, \delta_2 = 22^{\rm h}54^{\rm m}52\fs09, 
40\arcdeg54\arcmin33.20\arcsec$,
$\alpha_3, \delta_3 = 22^{\rm h}56^{\rm m}7\fs16, 
40\arcdeg45\arcmin22.04\arcsec$,
$\alpha_4, \delta_4 = 22^{\rm h}56^{\rm m}37\fs60, 
41\arcdeg3\arcmin29.06\arcsec$,
(J2000.0) \citep{ricker01b}.


\subsection{LOTIS Observations}
Although no coordinates were distributed with the real-time HETE-2
trigger, the $f/1.8$, 0.11-m LOTIS telescope \citep{park98a} observed
the position of the error box during routine sky patrol
on September~21 UT at 21.255~UT and 21.417~UT, only 52~min and 4.75 hours
after the GRB. The LOTIS telescope consists of four lens/camera
systems directed toward a common $8\fdg8~\times~8\fdg8$ field-of-view. 
Two of the LOTIS cameras are equipped with clear filters,
one with a standard Cousins $R$ filter, and one with the
standard Johnson $V$ filter. During sky patrol observations
the LOTIS telescope obtained 50~s exposures at each
position on the sky.

\subsection{SDSS PT Observations}
GRB010921 was also observed with the Sloan Digital Sky Survey's (SDSS;
York et al. 2000) Photometric Telescope (PT), which is located at
Apache Point Observatory (APO) in Sunspot, New Mexico. The PT is an
$f/8.8$, 0.5-m telescope equipped with $u'g'r'i'z'$ filters. The SDSS
is designed to be on the $u'g'r'i'z'$ photometric system described in
Fukugita et al. (1996) which is an $AB_\nu$ system where flat spectrum
objects ($F_\nu \propto \nu^0$) have zero colors. However, the current
photometric calibration of the PT may differ from this system by at
most a few percent (Stoughton et al. 2002), and we therefore denote
the PT photometric magnitudes by $u*g*r*i*z*$. The single SITe
$2048 \times 2048$ CCD camera has a $41.5\arcmin \times 41.5\arcmin$ 
field-of-view.
On September 22 UT beginning at 22.108 UT (21.33 hours after the GRB)
200 s exposures were taken in each of the $u'g'r'i'z'$ filters, and on
September 23 UT, beginning at 23.224 UT 200 s and 400 s exposures were
taken in each of the same filters (Lamb et al. 2001). Observations on
both nights covered the entire improved HETE-2 error box (Ricker et al.
2001, Hurley et al. 2001).  
Since the GRB exposures were unusually long, 
the afterglow was near the sky level in most of the images, and 
conditions were cloudy at the time, magnitude errors are 
uncharacteristically large.
 
\subsection{Super-LOTIS Observations}
The center of the IPN error box for GRB~010921 was
added to the Super-LOTIS sky patrol table and observations
of the location began shortly after nightfall on 
September~22.128~UT. Super-LOTIS is an $f/3.5$, 0.6-m Boller\&Chivens
telescope at Kitt Peak near Tucson, Arizona. Its focal plane array is a 
Loral $2048 \times 2048$ CCD camera covering $51\arcmin \times 51\arcmin$
field-of-view. The telescope system is fully automated and is capable of 
responding to a real-time GRB trigger within 30 seconds. A total
of twenty 50-second
exposures were obtained during the first epoch. Super-LOTIS
re-observed the field on September 22 and 23 UT beginning at 22.267, 
23.128, and 23.269 UT. Each of these observations also consisted of 
twenty 50-second exposures. For these observations, no astronomical
filter was used. 

\subsection{Tautenburg Observations}
On September~22.781 the Tautenburg Schmidt telescope began
observations of the error box in the Johnson $I$-band.
This telescope, located at Tautenburg, Germany,
is an f/2, 1.34-m aperture telescope equipped with a
SITe $2048 \times 2048$ CCD as the focal plane array. 
Its field-of-view is $36\arcmin \times 36\arcmin$.
Thirty-eight 120-second exposures were acquired.
The same field was re-observed on October~25 and eight 120-second 
exposures were acquired. During these exposures, the
telescope was dithered to allow a better treatment of bad pixels.

\subsection{USNOFS Host Galaxy Observations}
On November 17 UT (56 days after the burst), the USNOFS 1.0-m telescope 
at Flagstaff, Arizona, observed the GRB010921 
area to determine the brightness of the host galaxy in $Rc$ and $Ic$-bands. 
The USNOFS is an f/7.3 telescope equipped with $UBVRI$ filters; its
focal plane arrays are a SITe/Tektronix $2048 \times 2048$ CCD with 
$23\arcmin \times 23\arcmin$ field-of-view. 
For these observations, eight 
600-second exposures per filter were acquired and coadded.

\section{Results}

Early observations of the error box of GRB~010921 with
large aperture telescopes found no evidence of an
optical afterglow to the limit of the DPOSS
plates, $R \sim 20.5$ \citep{fox01a,henden01b}. However,
follow-up observations by \citet{price01a}
resulted in the report of a fading source which
displayed the characteristic decay behavior of
a GRB afterglow. The coordinates of the suspected afterglow are
$\alpha, \delta = 22^{\rm h}55^{\rm m}59.92\fs9,
+40\arcdeg55\arcmin52.83\arcsec$ (J2000.0). 

Following the reported detection of an optical afterglow 
candidate, we searched our images for the
optical afterglow. In the LOTIS clear- and $V$-band
images taken  52~min after the burst we found no
optical transient (OT) brighter than $m_{clear}$~$ >15.2 \pm 0.15$
and $V$~$ >15.6 \pm 0.15$.
The shutter for the $R$-band camera failed to open. In the second
epoch of LOTIS clear- and $V$-band images taken at September 21.417 UT 
(4.75 hours after the burst) we found no OT to the same 
limiting magnitude.

Analysis of the SDSS PT images obtained on September 22 UT reveals the
optical afterglow at $g* = 20.8 \pm 0.6, r* = 19.5 \pm 0.3$, and $i* =
18.8 \pm 0.7$, and gives $3\sigma$ upper limits of $u* > 20.5$ and $z*
> 15.0$. Analysis of the SDSS PT images obtained on September 23 UT
reveals the optical afterglow at $g* = 22.4 \pm 1.0$ and $r* = 21.5 \pm
1.0$, and gives $3\sigma$ upper limits of $u* > 19.5$, $i* > 18.5$, and
$z* > 18.0$. The completeness upper limits are the magnitudes at which
the probability of detection is 100\%. Because of variable extinction
over the field due to clouds, the 50\% detection limits are more than a
magnitude fainter than the completeness limits. The top panel of
Figure 1 ($a$ to $e$) shows the SDSS PT images in $u'g'r'i'z'$ 
obtained on September 22 UT of the GRB010921 afterglow area (marked by
a circle) reported by Price et al. (2001). 

\begin{figure}
\plotone{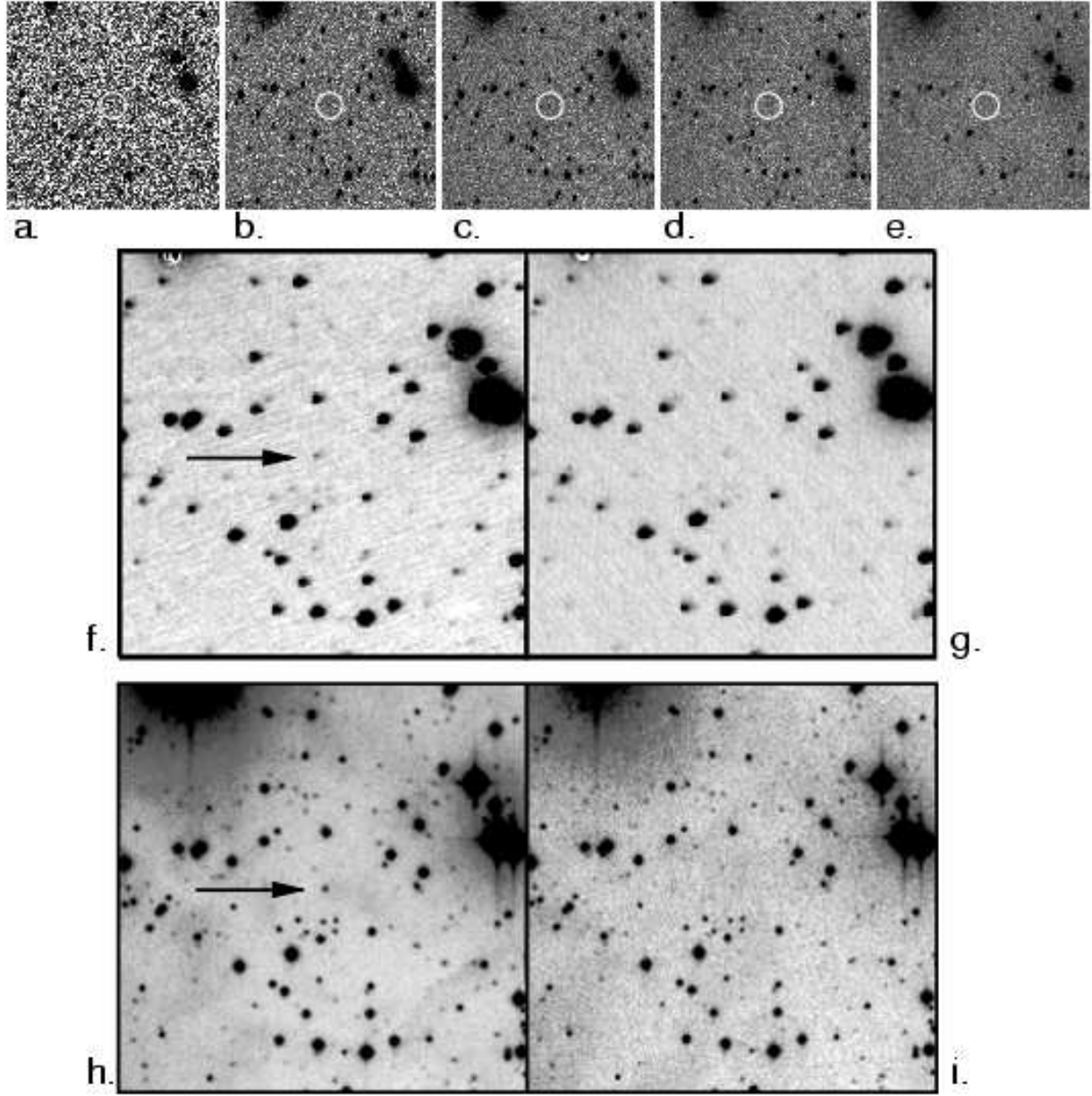}
\caption{Top Panels:  SDSS PT images of the afterglow of GRB010921 in
$u', g', r', i',$ and $z'$ (from left to right) taken on September 22 UT
beginning at 22.108 UT.  Middle Panels:  Super-LOTIS co-added images in
unflitered light taken on September 22 UT beginning at 22.128 UT (left
image) and at 22.267 (right image). Bottom Panels:  Tautenburg images
in I taken on September 22 UT (left image) and on October 25 (right
image). In all images, North is toward the top, and East is to the
left. The field-of-view of each of the images is $4' \times 4'$. The
location of the optical afterglow is indicated by circles in the SDSS
PT images, and by arrows in the Super-LOTIS and 
Tautenburg images. \label{fig:images}}
\end{figure}

We searched for the OT in the Super-LOTIS data by co-adding twenty images 
obtained during each epoch, two epochs per night. In the first co-added 
image, which was taken on September~22~UT beginning at 22.128~UT, we 
detect the optical afterglow at $m_{\rm clear}$~$=~19.4 \pm 0.2$ at 
15~$\sigma$ above noise level. The same analysis applied to the 
second epoch data (September~22.267~UT) detects the optical afterglow 
at $m_{\rm clear}$~$= 19.9 \pm 0.2$ at 12~$\sigma$ above the noise level.  
Panels $f$ and $g$ of Figure~\ref{fig:images} show the co-added images 
with the afterglow indicated by an arrow. We also searched for an 
OT in the data set obtained on September~23~UT and find no afterglow 
to the 10~$\sigma$ noise limit of $m_{\rm clear}$~$= 21.1 \pm 0.3$.

The Tautenburg data taken on September~22 UT and October~25 UT were 
co-added to search for the GRB~010921 afterglow. Panel $h$ of
Figure~\ref{fig:images} show the location of the
afterglow indicated by an arrow. Calibrating the afterglow
against secondary standards \citep{henden01a}, the
brightness of the afterglow is estimated as $I = 19.32 \pm 0.08$.
The host galaxy is marginally detected at the 5~$\sigma$ level in the 
second epoch image (October 25 UT at 25.770) at $I = 20.94 \pm 0.26$ 
(Panel $i$ of Figure~\ref{fig:images}). 

The USNOFS deep imaging data was analyzed to obtain the
brightness of the host galaxy. The host galaxy was clearly visible 
in the coadded image 56
days after the burst and we determine its brightness to be
$Rc = 21.93 \pm 0.09$ and $Ic = 21.05 \pm 0.08$.
 
\section{Implications and Conclusions}
\begin{figure}  
\plotone{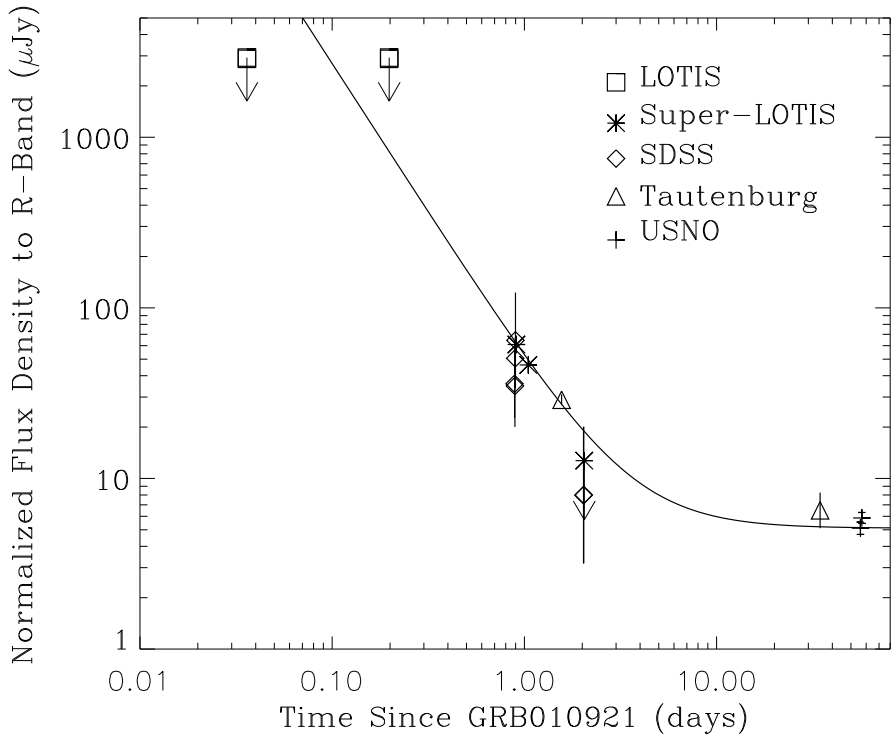}
\caption{The normalized flux density in $R$-band light curve of the 
afterglow of GRB010921, as constrained by the upper limits from 
the LOTIS telescope, and the measurements from the SDSS PT, the 
Super-LOTIS, the Tautenburg Schmidt, and the USNOFS telescopes. The 
fluxes in the different filters have been transformed to the R-band 
assuming a power-law spectrum of index 
$\beta = -2.3$ (Price et al. 2001). \label{fig:lightcurve}}
\end{figure}

Table 1 summarizes the magnitudes and upper limits in the various 
filters from all the above observations.
We transformed the fluxes measured in various filters to the $R$
filter by normalizing a $\beta=-2.3$ power-law spectrum 
\citep{price01a,kulkarni01a} 
to the effective wavelength and flux of each data point, and then
plotting the point at the effective wavelength of the $R$-band. This
calculated values are listed in the last column of Table 1.
Figure~\ref{fig:lightcurve} shows the resulting light curve of the
optical afterglow of GRB~010921. We fit the data 
with a power law decay plus a constant host galaxy 
flux, $F = F_0(t-t_0)^{-\alpha} + F_{host}$. 
We obtain a best fit decay index of $\alpha =1.75 \pm 0.28$
applying a host galaxy magnitude of $R = 21.93$ from the USNO 
measurements performed 56 days after the burst.
This is a typical value for an optical afterglow prior
to the jet break, which is predicted
to take place $\sim$ 130 days after the GRB based on the observed 
energetics of this burst \citep{djorg01a}.
The optical afterglow reported by \citet{price01a} was $R = 19.6
\pm 0.3$ on September 22 and significantly fain.ter on September 23 (we
place a limit on $R$ of roughly 20.5). 

We attempt to constrain the early-time power law decay
by extrapolating the best-fit power-law decay model back
to the LOTIS upper limit of
$R > 15.0$ on September~21.256. 
Figure~\ref{fig:lightcurve} shows that the early-time LOTIS
upper limits are inconsistent with an unchanging 
decay index from $t - t_0 = 52$~min to
$t - t_0 > 20$~h. This may suggest that the optical emission 
peaked at a magnitude fainter than the LOTIS limiting magnitude,
(perhaps similar to the afterglow of GRB 970508)
or that the slope changed between the observations, as suggested
in the case of GRB 991208 \citep{castro}. Complex light curve 
shapes at very early times have been observed (e.g., GRB970805)
and can be explained in terms of a distribution of Lorentz factors
produced by the central engine \citep{RandM} and the complex
evolution of multi-component shock emission in a relativistic fireball
\citep{kobayashi}. 

Our early time observations show that afterglow behavior can change
quickly within hours after the burst. With rapid localizations
we can probe the transition from the prompt emission phase to the 
subsequent unfolding of the canonical afterglow phase. GRB010921 only
provided an upper limit 52 minutes after outburst, but HETE-2 triggers
should eventually allow us to obtain simultaneous flux measurements.  
Robotic telescopes like LOTIS and Super-LOTIS are well suited to
finding this early-time emission in response to a near real-time
localization of the GRB by HETE-2.

\acknowledgments
Support for LOTIS and Super-LOTIS is provided 
by NASA (S-03975G and S-57797F) under the auspices 
of the U. S. Department of Energy by University of California
Lawrence Livermore National Laboratory (W-7405-Eng-48).
The Sloan Digital Sky Survey (SDSS) is a joint project of The
University of Chicago, Fermilab, the Institute for Advanced Study, the
Japan Participation Group, The Johns Hopkins University, the
Max-Planck-Institute for Astronomy (MPIA), the Max-Planck-Institute
for Astrophysics (MPA), New Mexico State University, Princeton
University, the U.S. Naval Observatory, and the University of
Washington. Apache Point Observatory, site of the SDSS telescopes, is
operated by the Astrophysical Research Consortium (ARC).
Funding has been provided by the Alfred P. Sloan
Foundation, the SDSS member institutions, NASA, the NSF, 
U.S. Department of Energy, Monbukogabusho, and the Max Planck Society.
K. Hurley is grateful for {\it Ulysses} support (JPL
958059) and for HETE support (MIT-SC-R-293291).


\clearpage

%

\begin{deluxetable}{ccccccc}
\tablecaption{Observations of GRB~010921. \label{tab:observations}}
\tablewidth{0pt}
\tablehead{
\colhead{Date}          & \colhead{$\Delta$T}    & \colhead{Telescope} & \colhead{Filter} &
\colhead{Exposure Time} & \colhead{Magnitude} & \colhead{$R$equivalent}  \\ 
\colhead{UT}            & \colhead{(days)}      & \colhead{}          & \colhead{}       &
\colhead{(s)}           & \colhead{}          & \colhead{Magnitude}
}
\startdata
Sep~21.219 & 0.000 & HETE-2      & GRB010921 &                   &                   &            \\ 
Sep~21.255 & 0.036 & LOTIS       & Clear     & $1 \times 50$~s   & $> 15.2 \pm 0.15$ & $>15.04$   \\ 
Sep~21.255 & 0.036 & LOTIS       & $V$       & $1 \times 50$~s   & $> 15.6 \pm 0.15$ & $>15.05$   \\ 
Sep~21.417 & 0.198 & LOTIS       & Clear     & $1 \times 50$~s   & $> 15.2 \pm 0.15$ & $>15.04$   \\ 
Sep~21.255 & 0.198 & LOTIS       & $V$       & $1 \times 50$~s   & $> 15.6 \pm 0.15$ & $>15.05$   \\ 
Sep~22.108 & 0.889 & SDSS PT     & $u*$      & $500$~s           & $> 20.5 \pm 0.5$  & $>19.82$   \\ 
Sep~22.112 & 0.893 & SDSS PT     & $g*$      & $200$~s           & $20.8 \pm 0.6$    & 19.85      \\ 
Sep~22.115 & 0.896 & SDSS PT     & $r*$      & $200$~s           & $19.5 \pm 0.3$    & 19.44      \\ 
Sep~22.118 & 0.899 & SDSS PT     & $i*$      & $200$~s           & $18.8 \pm 0.7$    & 19.49      \\ 
Sep~22.121 & 0.902 & SDSS PT     & $z*$      & $200$~s           & $> 15.0 \pm 0.5$  & $>15.39$   \\ 
Sep~22.128 & 0.909 & Super-LOTIS & Clear     & $20 \times 50$~s  & $19.4 \pm 0.1$    & 19.24      \\ 
Sep~22.267 & 1.048 & Super-LOTIS & Clear     & $20 \times 50$~s  & $19.9 \pm 0.1$    & 19.54      \\ 
Sep~22.781 & 1.562 & Tautenburg  & $Ic$      & $38 \times 120$~s & $19.32 \pm 0.08$  & 20.05      \\ 
Sep~23.128 & 1.909 & Super-LOTIS & Clear     & $20 \times 50$~s  & $> 21.1 \pm 0.3$  & $>20.94$   \\ 
Sep~23.244 & 2.025 & SDSS PT     & $u*$      & $500$~s           & $> 19.5 \pm 0.5$  & $>18.82$   \\ 
Sep~23.249 & 2.030 & SDSS PT     & $g*$      & $400$~s           & $  22.4 \pm 1.0$  & 21.45   \\ 
Sep~23.255 & 2.036 & SDSS PT     & $r*$      & $400$~s           & $  21.5 \pm 1.0$  & 21.44   \\ 
Sep~23.260 & 2.041 & SDSS PT     & $i*$      & $400$~s           & $> 18.5 \pm 0.5$  & $>19.19$   \\ 
Sep~23.266 & 2.047 & SDSS PT     & $z*$      & $400$~s           & $> 18.0 \pm 0.5$  & $>19.18$   \\ 
Sep~23.269 & 2.050 & Super-LOTIS & Clear     & $20 \times 50$~s  & $> 21.2 \pm 0.3$  & $>20.94$   \\ 
Oct~25.770 & 34.551 & Tautenburg & $Ic$      & $8 \times 120$~s  & $20.94 \pm 0.26$  & 21.67   \\ 
Nov~17.7   & 56.700 & USNOFS     & $Rc$      & $8 \times 600$~s  & $21.93 \pm 0.09$  & 21.93   \\ 
Nov~17.7   & 56.700 & USNOFS     & $Ic$      & $8 \times 600$~s  & $21.05 \pm 0.08$  & 21.78   \\ 
\enddata
\end{deluxetable}
\clearpage

%
\clearpage

\end{document}